\documentclass[twocolumn,superscriptaddress]{revtex4}
\usepackage[dvips]{graphicx}
\usepackage{times}
\usepackage{amsmath,amssymb}
\begin{document}
\title{%
Bunching visibility for 
correlated photons from single GaAs quantum dots
}
\author{T.~Kuroda}
\homepage[URL: ]{http://www.nims.go.jp/laser_kuroda/}
\affiliation{%
National Institute for Materials Science, 
1-1 Namiki, Tsukuba 305-0044, Japan}
\author{T.~Belhadj}
\affiliation{%
National Institute for Materials Science, 
1-1 Namiki, Tsukuba 305-0044, Japan}
\affiliation{%
Universit\'{e} de Toulouse, LPCNO, INSA-CNRS-UPS, 
135 avenue de Rangueil, 31077 Toulouse Cedex 4, France}
\author{M.~Abbarchi}
\author{C.~Mastrandrea}
\author{M.~Gurioli}
\affiliation{%
LENS, and Dipartimento di Fisica, Universit\`{a} di Firenze, 
Via Sansone 1, I-50019, Sesto Fiorentino, Italy}
\author{T.~Mano}
\author{N.~Ikeda}
\author{Y.~Sugimoto}
\author{K.~Asakawa}
\author{N.~Koguchi}
\altaffiliation[Present address: ] {%
L-NESS and Dipartimento di Scienza dei Materiali, 
Universit\'{a} di Milano--Bicocca, Via Cozzi 53, I-20125, Milano, Italy.}
\author{K.~Sakoda}
\affiliation{%
National Institute for Materials Science, 
1-1 Namiki, Tsukuba 305-0044, Japan}
\author{B. Urbaszek}
\author{T. Amand}
\author{X. Marie}
\affiliation{%
Universit\'{e} de Toulouse, LPCNO, INSA-CNRS-UPS, 
135 avenue de Rangueil, 31077 Toulouse Cedex 4, France}
\date{\today}
\begin{abstract}
We study photon bunching phenomena associated with biexciton-exciton cascade in single GaAs self-assembled quantum dots. Experiments carried out with a pulsed excitation source show that significant bunching is only detectable at very low excitation, where the typical intensity of photon streams is less than the half of their saturation value. Our findings are qualitatively understood with a model which accounts for Poissonian statistics in the number of excitons, predicting the height of a bunching peak being determined by the inverse of probability of finding more than one exciton. 
\end{abstract}
\maketitle
\section{Introduction}
Semiconductor quantum dots (QDs) are 
nonconventional light emitters which produce nonclassical photons. 
Since the first demonstration of photon antibunching in single QD photoluminescence (PL) \cite{Michler_Nat00, Lounis_CPL00}, 
numerous attempts have been made to develop efficient single-photon sources 
based on QDs embedded in micro- and nanostructures \cite{MKB00,SPS01,ZBJ01,MRG01,YKS02}. 

The nonclassical nature of photons has commonly been verified by measuring coincidence probability 
for the arrival of two photons, utilizing a Humbury, Brown and Twiss setup. 
In the case of single photons, the coincidence histogram shows an antibunching dip (negative peak), 
confirming zero probability of finding two photons at the same time. 
The depth of the histogram represents a figure of merit for a single photon source. 
For correlated photons, on the other hand, the histogram would present a bunching peak, suggesting a higher probability of finding a particular pair of photons. 

Two photons sequentially emitted with a biexciton--exciton cascade are an example of the correlated photons \cite{MRM01,RMG01,KFB02}, and are recognized as a resource of entanglement \cite{BSP00}. Polarization entangled photons in this framework were demonstrated in bulk CuCl \cite{Eda_Nat04}; subsequently the researchers improved the visibility of coincidence, reaching violation of Bell's inequality, by adopting high repetition pulses which engender suppression of `\textit{accidental coincidence counts}' \cite{Oohata_PRL07,Eda_JJAP}.
Triggered entangled photons have recently been demonstrated in semiconductor QDs \cite{SYA06,ALP06}. Once again, achieving high fidelity has becomed a central issue \cite{Young_NJP06,Hafenbrak_NJP07}, and coincidence data are often analyzed after making `\textit{background subtraction}'. These studies imply that weak excitation yields high fidelity, i.e., high bunching peaks with low backgrounds, while the physical origin has not been clarified. 

\begin{figure}
\includegraphics[scale=0.7]{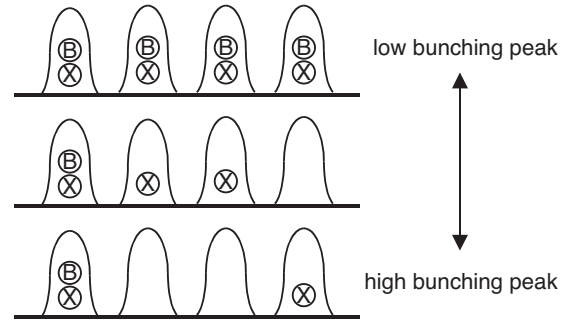}
\caption{%
Schematic representation of the observation of high bunching peaks for correlated biexcitonic (B) and excitonic (X) photons.
}
\label{fig}
\end{figure}

In this work we are studying the excitation power dependence of photon bunching associated with biexciton--exciton cascades. We have found that a significant bunching feature is only present at low excitation, and disappears at high excitation where the intensity of biexciton/exciton PL approaches its saturation level. We demonstrate that a smaller visibility at higher excitation is not due to normal high excitation effects, such as the onset of emissions from multiple carrier states, a wetting layer, or incoherent carrier scatterings. Instead, our findings are understood by a purely photon-statistical effect, where a bunching peak decreases with increasing probability of finding each photon in a pulse, as is illustrated in Fig.~\ref{fig}. This is quite surprising because a regulated sequence of correlated two photons does not show any bunching feature, although such a light source should be favorable for practical applications.

\section{Experimental procedure}
The experiments were performed on GaAs self-assembled QDs in an Al$_x$Ga$_{1-x}$As ($x=0.26 \pm0.01$) barrier grown on a (311)A surface by droplet epitaxy \cite{KTC91,Watanabe1}. Atomic force microscopy demonstrated the formation of lens-shaped QDs of 80~nm in diameter, 10~nm in height, and $5\times10^{9}$~cm$^{-2}$ in density. 
The QDs were embedded in a two-dimensional photonic crystal (PhC) membrane, where air holes normal to the surface were regularly made with $C_{6v}$ symmetry. In the center of PhC, three missing holes were arranged in line, forming L3 defect cavity. The PhC membranes of 140~nm in thickness, 204~nm in lattice constant, and 80~nm in the air-hole diameter, were fabricated by electron beam lithography and reactive ion beam etching \cite{sugimoto, ikeda}. Spectral characteristics in the PhC membranes are presented elsewhere \cite{Kuroda_APL08}. 

For the optical study, we used second-harmonic output of an optical parametric oscillator synchronously pumped by a mode-locked Ti-sapphire laser. The laser system produced excitation pulses with 3~ps duration and 76~MHz repetition. The wavelength was tuned to be 640~nm, exciting the absorption edge of the AlGaAs barrier. 

\begin{figure}
\includegraphics[scale=0.7]{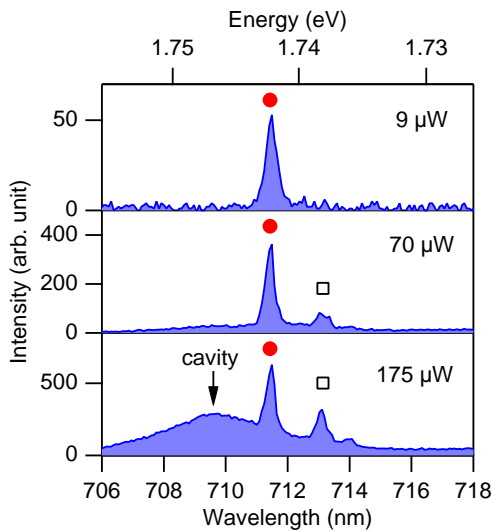}
\caption{%
(color online) A series of microPL spectra for a single GaAs QD at various excitation intensities. 
Photon energies for excitonic and biexcitonic lines are presented by $\bullet$ and $\Box$, respectively.}
\label{fig_spctr}
\end{figure}

PL from a single GaAs QD was observed with a confocal microPL setup, using an objective lens of 4~mm focal length and 0.42 in numerical aperture. The PL beam was split by a 1:1 beam splitter, with each beam fed in a grating spectrometer equipped with a fiber-coupled avalanche photodiode (APD). The spectral window for each APD was around 0.8~meV, being much smaller than the exciton--biexciton split in a PL spectrum (see Fig.~\ref{fig_spctr}). Then, we set the detection wavelength of one APD to the exciton line, and  that of another APD to the biexciton line. Electric output from APD's was sent to a coincidence counter, yielding a start/stop event for the photon arrival. PL spectra were also monitored by a charge coupled device. All experiments were performed at 8~K. 

\section{Experimental results and discussions}
Figure~\ref{fig_spctr} presents the emission spectra of the QD which we will be examining in this study. 
At low excitation, a single emission line ($\bullet$) is present at 711.5~nm (1.742~eV), 
being assigned to PL from neutral excitons (X). With increasing excitation intensity, 
another emission line ($\Box$) starts to appear at 713.1~nm (1.738~eV), at the lower energy side of the X line, which is assigned to PL from biexcitons (B). The split from the X line to the B line is 3.9~meV, which is typical for GaAs QDs grown by droplet epitaxy \cite{Kuroda_PRB,Kuroda_APL,abbarchi}. 

At very high excitation, a broad spectral component emerges at 709~nm (1.747~eV) with 5~meV in full width at half maximum. The similar behavior is observed in PhC samples at high excitation. It has also been found that the energy depends systematically on the lattice constant of PhC \cite{Kuroda_APL08}. We therefore attribute this emission to the highly-excited PL continuum enhanced by the cavity resonance. Note that we choose a PhC cavity with a relatively low quality factor ($Q\sim300$), allowing both B and X lines to be on resonance, while retaining their high emission efficiency. For the coincidence study, we will limit excitation power to below the value corresponding to the onset of the cavity emission. 

\begin{figure}
\includegraphics[scale=0.7]{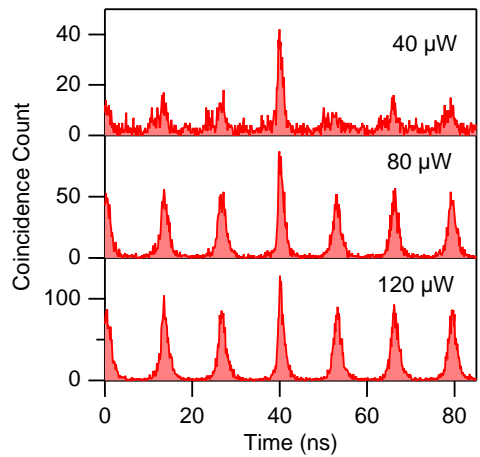}
\caption{%
(color online) Histograms for the coincident arrival of two photons with biexcitonic and excitonic transitions with 164~ps time bin for three excitation intensities. The integration times were 90, 30, and 15 minutes for the top, middle, and bottom panels, respectively. 
}
\label{data}
\end{figure}

The autocorrelation trace of the X line (not shown) presents an antibunching dip at coinciding times ($\sim39.9$~ns in Fig.~\ref{data}), demonstrating its single-photon behavior. 
Coincidence histograms for cross-correlations between the B and X photons are presented in Fig.~\ref{data}. At low excitation with 40~$\mu$W, the histogram shows a high central peak following sequential backgrounds. The presence of a high coincidence peak confirms that two photons were generated with a single radiative cascade. The sequential background is due to coincidence counts between photons in temporally separated pulses, which were emitted synchronously with the laser source. The bunching visibility, i.e., the ratio of a central coincidence peak to background side peaks, is evaluated to be 2.7 ($\pm 0.1$). 

When excitation intensity increases to 80~$\mu$W, we find a remarkable reduction in the relative height of a bunching peak, while the histogram shows a higher signal-to-noise ratio, reflecting greater counting rates. The visibility in this case is evaluated to be 1.3 ($\pm 0.05$). For excitation intensity at 120~$\mu$W, the coincidence peak further decreases to 1.06 ($\pm 0.02$), as shown in the bottom panel of Fig.~\ref{data}. 
We should note that negligible bunching peak was also observed in our previous work, which studied QDs 
without PhC processing \cite{Kuroda_APEX}, suggesting that this is a general feature related to the high excitation regime. 

We are analyzing the power dependence of coincidence histograms in terms of photon number statistics: Preliminary theoretical work was reported in \cite{arxiv}. In this model, we assume that the probability of finding B and X photons is simply determined by the number of photoinjected excitons, obeying the following Poissonian distribution function, 
\begin{equation}
P_{\bar{N}}(n)=\exp(-\bar{N})\frac{\bar{N}^n}{n!}, 
\label{poisson}
\end{equation}
where $\bar{N}$ is the mean number of excitons. 
Since an X photon is generated when the number of excitons is more than one, and both B and X photons are generated when more than two excitons are present initially, the probability of finding an X (B) photon in a pulse, $P_X$ ($P_B$), is given by; 
\begin{align}
P_X &= \sum_{n\ge1}P_{\bar{N}}(n) =1-\exp(-\bar{N}), \label{eqX} \\ 
P_B &= \sum_{n\ge2}P_{\bar{N}}(n) =1-\exp(-\bar{N})(1+\bar{N}), \label{eqB}
\end{align}
where we used a relation, $\sum_{n\ge0}P(n)=1$. 

A coincidence peak, $g^{(2)}(0)$, is given by a joint probability for both $B$ and $X$ photons being present in a pulse, and of being counted by two detectors, whose counting yield is $\eta_B$ and $\eta_X$, respectively. Thus we find, 
\begin{equation}
g^{(2)}(0) =  \sum_{n\ge2}P(n)\eta_B\eta_X = P_B\eta_B\eta_X.
\end{equation}
Note that the parameter of $\eta_B$ and $\eta_X$ accounts for every effect which causes a counting loss, including a finite efficiency of photon extraction and photon detection.

A coincidence background, $g^{(2)}_{BG}$, is written by a product of counting probability for B and X photons, each belonging to an uncorrelated pulse. Thus, we find, 
\begin{equation}
g^{(2)}_{BG} =  P_B\eta_BP_X\eta_X.
\end{equation}
%
\begin{figure}
\includegraphics[scale=0.7]{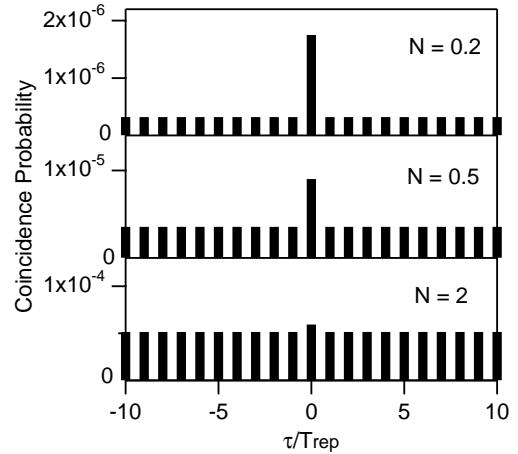}
\caption{%
Coincidence probability for correlated photon pairs for various values of mean exciton number, $N$, for $\eta_B=\eta_X=10^{-2}$. The horizontal (temporal) axis is scaled by pump repetition, $T_d$.}
\label{fig_3}
\end{figure}

Figure~\ref{fig_3} illustrates expected coincidence histograms for various values of $\bar{N}$. It shows a central bunching peak following constant side peaks. The relative height of the central peak is found to depend on $\bar{N}$, and it is larger for smaller numbers of $\bar{N}$. 
The bunching visibility is, therefore, expressed by, 
\begin{equation}
g^{(2)}(0)/g^{(2)}_{BG}= P_X^{-1}= \{1-\exp(-\bar{N})\}^{-1}.
\label{eq_vsblt}
\end{equation}
The above equation suggests that a bunching peak becomes higher as the probability of finding a dark pulse emitted from ``zero" excitons increases, as is schematically shown in Fig.~\ref{fig}. 
Note that the above simple expression is also obtained by rigorous formulation based on the quantum regression theorem, as is developed in Appendix. 

\begin{figure}
\includegraphics[scale=0.7]{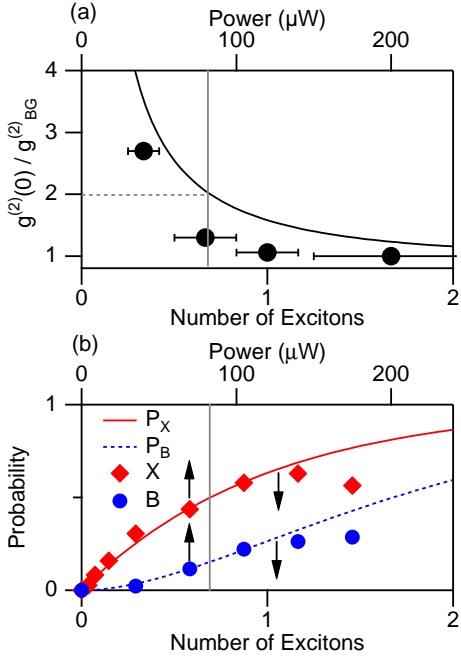}
\caption{%
(color online) (a) Relative height of bunching peaks, $g^{(2)}(0)/g^{(2)}_{BG}$, as a function of mean numbers of excitons. The solid circles show the bunching visibilities that are obtained experimentally, and the solid line shows the theoretical dependence which follows Eq.~\ref{eq_vsblt} . For comparison, the same power dependence of the intensity of exciton and biexciton emissions is plotted in (b).
}
\label{fidelity}
\end{figure}
 
The bunching visibility sharply decreases as the mean number of excitons increases, as shown by the solid line in Fig.~\ref{fidelity}(a). The same dependence of probability of finding B and X photons, which is proportional to the relevant PL intensity, is presented in Fig.~\ref{fidelity}(b). Through the comparison between Figs.~\ref{fidelity}(a) and \ref{fidelity}(b), we find that a very low injection of excitons is necessary to realize a significant bunching peak. As a guideline, we plot a vertical line which intersects the visibility curve at $g^{(2)}(0)/g^{(2)}_{BG}=2$, the value being available with classical incoherent light. It is shown that to achieve high bunching visibility beyond classical criteria, we should keep the number of excitons lower than $\ln2 =0.69$. 
Note that this number is much smaller than two, which is a rough index for a biexciton to be present in a QD. Correspondingly, the intensity of the B (X) line should be less than 0.153 (0.5) of its saturation intensity.  

Let us compare the above theoretical prediction with experimental data. For simplicity, we assume the number of excitons being proportional to the excitation intensity. Then, we evaluate the efficiency of photoinjection through the fit of the X and B intensities to the power dependence of Eqs.~\ref{eqX} and \ref{eqB}, respectively. The result of this fit is plotted by diamonds (X) and circles (B) in Fig.~\ref{fidelity}(b). These PL intensities agree with this model. With the use of this injection efficiency, we can estimate the number of excitons for each coincidence histogram. The observed data for the coincidence visibility as a function of exciton number are finally plotted by solid circles in Fig.~\ref{fidelity}(a). 
We find that the experimental trend, i.e., a steep reduction in bunching visibility is reproduced quite well by this model, while the observed visibility is significantly smaller than the theoretical ones. 

One likely reason for this discrepancy is the influence of emission from multiple carrier states (or a wetting layer). The onset of these incoherent signals results in the emission of uncorrelated photons, thus the reduction of bunching visibility. Moreover, in the above treatment, we assumed that excitons followed Poissonian statistics, and their mean number was proportional to the excitation power. However, neither hypothesis is, in fact, relevant at high excitation when the exciton number approaches the maximum number of carriers which can be occupied by a QD. At such high excitation both B and X photons are likely more saturated than in this model, thus the actual visibility becomes lower than that of Fig.~\ref{fidelity}(a). 

\section{Conclusions}
In conclusion, 
we have demonstrated the fidelity of bunching statistics associated with biexciton-exciton cascades depends on excitation intensity, and significant bunching feature only appears at very low excitation such that  $\bar{N}\ll2$. This situation is in stark contrast to the experimental condition for characterizing a single-photon emitter, where an antibunching dip is normally measured at sufficiently high excitation, realizing regulated single-photon pulses, and high counting rates. On the other hand, the usage of \textit{dilute} photon pulses is essential for characterizing the bunching feature for correlated photons, although it is often difficult to take coincidence measurements at sufficiently low excitation. 

In atomic physics a model for a three-level system under low continuous-wave excitation describing the power dependence of photon bunching has been derived \cite{Loudon}.
Here, we have analyzed photon statistics between biexcitonic and excitonic recombinations with pulsed excitation. Note that the similar power sensitive behavior should be involved in polarization-resolved cross-correlation experiments. In this case the reduction of coincidence fidelity is induced by several microscopic effects such as a fine level split of excitons, spin relaxation, and the onset of unwanted emissions, as well as the photon saturation effect which we have studied. 

\begin{acknowledgments}
We are grateful to Keiji Kuroda, Tetsuyuki Ochiai, Shunsuke Ohkouchi, and Professor Fujio Minami for their valuable discussions. This work was partially supported by a Grant-in-Aid from the Japan Society for the Promotion of Science (JSPS). T.~K. gratefully acknowledges support of PRESTO, Quanta and Information, from Japan Science and Technology Agency (JST). 
\end{acknowledgments}

\appendix*
\section{Second-order correlation function for sequential pulsed excitation}
\begin{figure}[t]
\includegraphics[scale=0.7]{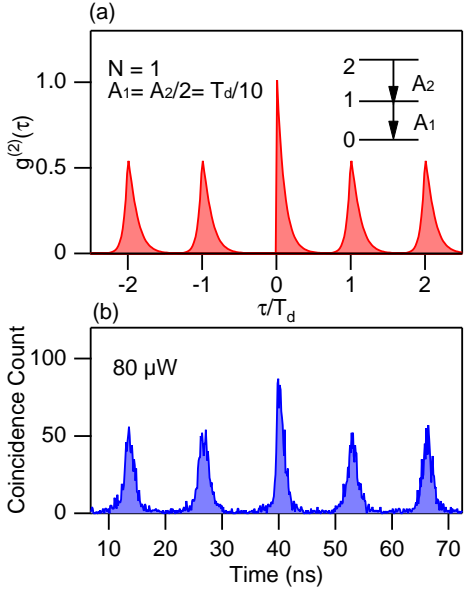}
\caption{%
(color online) (a) Calculated second-order correlation function for $A_1 = A_2/2 = T_d/10$, and Poissonian distributed exciton population with $\bar{N}=1$, which corresponds to $\rho_{22}(0)\simeq 0.264$ and $\rho_{11}(0) \simeq 0.6321$, according to Eqs. 2 and 3. (b) Measured coincidence histogram at 80~$\mu$W.
}
\label{crrltn}
\end{figure}

We will be deriving the expression of the interbeam second-order correlation function associated with biexciton-exciton cascades. Let us assume that QDs are excited by short optical pulses, and the pulse interval, $T_d$, is sufficiently longer than the relaxation times of QDs. We restrict ourselves to the three levels consisting of the biexciton $|2\rangle$, exciton $|1\rangle$, and vacuum $|0\rangle$ states, as depicted in the inset of Fig.~\ref{crrltn}. The presence of higher excited states influences the initial population of $|2\rangle$, followed by the fast energy relaxation after initial photoinjection. The population dynamics is characterized by a set of rate equations, 
\begin{gather}
d\rho_{22}(t)/dt=-A_2\rho_{22}(t), \\
d\rho_{11}(t)/dt=A_2\rho_{22}(t)-A_1\rho_{11}(t).
\label{rate_eq}
\end{gather}
where $\rho_{ii}$ denotes the diagonal density matrix element of the $i$ exciton level, and $A_2$ ($A_1$) is a biexciton--exciton (exciton--vacuum) transition rate. The above equations have the general solutions, 
\begin{gather}
\rho_{22}(t)=\rho_{22}(0)e^{-A_2t}, \\
\rho_{11}(t)=-\frac{A_2\rho_{22}(0)}{A_2-A_1}e^{-A_2t}+
\{\rho_{11}(0)+\frac{A_2\rho_{22}(0)}{A_2-A_1}\}e^{-A_1t}.
\end{gather}

The second-order correlation function is presented in terms of dipole projection operators, $\pi_2=|1\rangle \langle 2|$, $\pi_1=|1\rangle \langle 0|$, and their conjugates  \cite{Loudon}, 
\begin{equation}
g^{(2)}_{21}(t, t+\tau)=\frac{
\langle \pi_2^{\dagger}(t)\pi_1^{\dagger}(t+\tau)\pi_1(t+\tau)\pi_2(t) \rangle}
{\langle \pi_2^{\dagger}(t)\pi_2(t) \rangle \langle \pi_1^{\dagger}(t+\tau)\pi_1(t+\tau) \rangle},
\label{crrltn1}
\end{equation}
The quantum regression theorem allows to express double-time expectation values in the right hand side of Eq.~\ref{crrltn1}, in terms of single expectation values. Thus, we obtain, 
\begin{equation}
g^{(2)}_{21}(t, t+\tau)=\frac{
\langle \pi_2^{\dagger}(t)\pi_2(t)\rangle e^{-A_1\tau}}
{\langle \pi_2^{\dagger}(t)\pi_2(t) \rangle \langle \pi_1^{\dagger}(t+\tau)\pi_1(t+\tau) \rangle}. 
\end{equation}

In the experiments, the coincidence counts are acquired over a long integration time, $T_{int} (\gg T_d)$. The time-integrated coincidence counts are, therefore, given by, 
\begin{align}
G^{(2)}(\tau) &=\int_0^{T_{int}} g^{(2)}_{21}(t, t+\tau) dt, \\
&\approx e^{-A_1\tau}I_2/I_1I_2, \\
&=\frac{A_1e^{-A_1\tau}}{\rho_{11}(0)+\rho_{22}(0)} \label{G2_0}
\end{align}
where
\begin{align}
I_1
&=\int_0^{\infty} \langle \pi_1^{\dagger}(t)\pi_1(t) \rangle dt 
=\frac{\rho_{11}(0)+\rho_{22}(0)}{A_1}. \\
I_2
&=\int_0^{\infty}\langle \pi_2^{\dagger}(t)\pi_2(t)\rangle dt
=\frac{\rho_{22}(0)}{A_2}. 
\end{align}

Note that the value of $I_1$ ($I_2$) is proportional to the averaged intensity of the biexciton--exciton (exciton--vacuum) transition. Equation~\ref{G2_0} presents that $G^{(2)}(\tau) =0$ for $\tau<0$, and it decays with the exciton decay rate for $\tau>0$. In addition, the shape of $G^{(2)}(\tau)$ is independent of the initial population. 

Next, we consider photon coincidence events between uncorrelated pulses, which cause background side peaks in coincidence histograms. In this case, double-time expectation values in Eq.~\ref{crrltn1} are decomposed in the product of two expectation values; 
\begin{gather}
\langle \pi_2^{\dagger}(t)\pi_1^{\dagger}(t+T_d+\tau)\pi_1(t+T_d+\tau)\pi_2(t) \rangle \notag\\
=\langle \pi_2^{\dagger}(t) \pi_2(t) \rangle \langle\pi_1^{\dagger}(t+\tau)\pi_1(t+\tau)\rangle.
\end{gather}
Thus, we obtain the time-integrated coincidence counts given by, 
\begin{align}
G_{BG}^{(2)}&(\tau+T_d) \notag\\
&=\frac{1}{I_1I_2}\int_0^{\infty}\langle \pi_2^{\dagger}(t) \pi_2(t) \rangle \langle\pi_1^{\dagger}(t+\tau) \pi_1(t+\tau)\rangle dt  \notag\\
&=
\begin{cases}
	 \frac{1}{I_1I_2} (I_3-I_4)e^{A_2\tau} &(\tau\leq0), \\
	 \frac{1}{I_1I_2} (I_3e^{-A_1\tau}-I_4e^{-A_2\tau}) &(\tau\geq0), \label{G2_BG}
\end{cases}
\end{align}
where 
\begin{gather}
I_3=\frac{
		(A_2-A_1)\rho_{11}(0)\rho_{22}(0)+A_2\rho_{22}(0)^2
		}{
		A_2^2-A_1^2	
		}, \\
I_4=\frac{\rho_{22}(0)^2}{2(A_2-A_1)}.
\end{gather}

Equation~\ref{G2_BG} shows that each side peak rises with the biexciton decay rate, and it decays dominantly with the exciton decay rate. Calculation results for the second-order correlation functions are presented in Fig.~6(a), together with an experimental coincidence curve in Fig.~6(b). Asymmetric shape in both central and side peak is well reproduced by the simulation. However, we can only deal with the area of $G^{(2)}(\tau)$ for quantitative analysis, because of the fast recombination of GaAs QDs, whose time scale is similar with that of the instrumental response function. In this case, bunching visibility defined by Eq.~\ref{eq_vsblt} becomes, 
\begin{equation}
\frac{\int _{-\infty}^{\infty}G^{(2)}(\tau)d\tau}{\int_{-\infty}^{\infty} G_{BG}^{(2)}(\tau+T_d)d\tau}=\frac{1}{\rho_{11}(0)+\rho_{22}(0)} \label{vsblt1}
\end{equation}
Note that Eq.~\ref{vsblt1} is equivalent to Eq.~\ref{eq_vsblt}, because the value of $\{\rho_{11}(0)+\rho_{22}(0)\}$ is nothing but the probability of finding more than one exciton in a QD. 
%


\begin{thebibliography}{99}
\bibitem{Michler_Nat00}
	P.~Michler, A.~Imamo\u{g}lu, M.~D.~Mason, P.~J.~Carson, G.~F.~Strouse, and S.~K.~Buratto, 
	Nature \textbf{406}, 968 (2000). 
	
\bibitem{Lounis_CPL00}
	B.~Lounis, H.~A.~Bechtel, D.~Gerion, P.~Alivisatos, and W.~E.~Moerner, 
	Chem. Phys. Lett. \textbf{329}, 399 (2000). 

\bibitem{MKB00}
	P. Michler, A. Kiraz, C. Becher, W. V. Schoenfeld, P. M. Petroff, L. Zhang, 
	E.~Hu, and A.~Imamo\u{g}lu,  
	Science \textbf{290}, 2282 (2000).
	
\bibitem{SPS01}
	C.~Santori, M.~Pelton, G.~Solomon, Y.~Dale, and Y.~Yamamoto, 
	Phys. Rev. Lett. \textbf{86}, 1502 (2001). 
	
\bibitem{ZBJ01}
	V.~Zwiller, H.~Blom, P.~Jonsson, N.~Panev, S.~Jeppesen, T.~Tsegaye, 
	E.~Goobar, M.~Pistol, L.~Samuelson, and G.~Bj\"{o}rk, 
	Appl. Phys. Lett. \textbf{78}, 2476 (2001). 
	
\bibitem{MRG01}
	E.~Moreau, I.~Robert, J.-M.~G\'{e}rard, I.~Abram, L.~Manin, and V.~Thierry-Mieg, 
	Appl. Phys. Lett. \textbf{79}, 2865 (2001).

\bibitem{YKS02}
	Z.~Yuan, B.~E.~Kardyanal, R.~M.~Stevenson, A.~J.~Shields, C.~J.~Lobo, 
	K.~Cooper, N.~S.~Beattie, D.~A.~Ritchie, and M. Pepper, 
	Science \textbf{295}, 102 (2002).


\bibitem{MRM01}
	E. Moreau, I. Robert, L. Manin, V. Thierry-Mieg, J.-M. G\'{e}rard, 
	and I. Abram, 
	Phys. Rev. Lett. \textbf{87}, 183601 (2001). 

\bibitem{RMG01}
	D.~V.~Regelman, U. Mizrahi, D. Gershoni, E. Ehrenfreund, W.~V.~Schoenfeld, 
	and P. M. Petroff, 
	Phys. Rev. Lett. \textbf{87}, 257401 (2001).

\bibitem{KFB02}
	A. Kiraz, S. F\"{a}lth, C. Becher, B. Gayral, W. V. Schoenfeld, 
	P. M. Petroff, L. Zhang, E. Hu, and A. Imamo\u{g}lu, 
	Phys. Rev. B \textbf{65}, 161303 (2002).

\bibitem{BSP00}
	O.~Benson, C.~Santori, M. Pelton, and Y. Yamamoto, 
	Phys. Rev. Lett. \textbf{84}, 2513 (2000).

\bibitem{Eda_Nat04}
	K. Edamatsu, G. Oohata, R. Shimizu, and T. Itoh, 
	Nature \textbf{431}, 167 (2004). 

\bibitem{Oohata_PRL07}
	G. Oohata, R. Shimizu, and K. Edamatsu, 
	Phys. Rev. Lett. \textbf{98}, 140503 (2007). 

\bibitem{Eda_JJAP}
	K. Edamatsu, Jpn. J. Appl. Phys. \textbf{46}, 7175 (2007).
	
\bibitem{SYA06}
	R.~M.~Stevenson, R.~J.~Young, P.~Atkinson, D.~A.~Ritchie, and A.~J.~Shields, 
	Nature \textbf{439}, 179 (2006).
	
\bibitem{ALP06}
	N. Akopian, N. H. Lindner, E. Poem, Y. Berlatzky, J. Avron, D. Gershoni, 
	B. D. Gerardot, and P.~M.~Petroff, 
	Phys. Rev. Lett. \textbf{96}, 130501 (2006).
	
\bibitem{Young_NJP06}
	R.~J.~Young, R. M. Stevenson, P. Atkinson, K. Cooper, D. A. Ritchie, and A. J. Shields, 
	New J. Phys. \textbf{8}, 29 (2006). 

\bibitem{Hafenbrak_NJP07}
	R. Hafenbrak, S. M. Ulrich, P. Michler, L. Wang, A. Rastelli, and O. G. Schmidt, 
	New J. Phys. \textbf{9}, 315 (2007). 

\bibitem{KTC91}
    N. Koguchi, S. Takahashi, and T. Chikyow,
    J. Cryst. Growth \textbf{111}, 688 (1991).

\bibitem{Watanabe1}
    K. Watanabe, N. Koguchi, and Y. Gotoh,
    Jpn. J. Appl. Phys. Part 2, \textbf{39}, L79(2000).

\bibitem{sugimoto}
	Y. Sugimoto, N. Ikeda, N. Carlsson, K. Asakawa, H. Kawai, and K. Inoue, 
	J. Appl. Phys. \textbf{91}, 922 (2002). 
    
\bibitem{ikeda}
	N. Ikeda, Y. Sugimoto, Y. Tanaka, K. Inoue, and K. Asakawa, 
	IEEE J. Sel. Area Comm. \textbf{23}, 1315 (2005). 

\bibitem{Kuroda_APL08}
	T.~Kuroda, N. Ikeda, T. Mano, Y. Sugimoto, T. Ochiai, K. Kuroda, 
	S. Ohkouchi, N. Koguchi, K. Sakoda, and K. Asakawa, 
	Appl. Phys. Lett., \textbf{93}, 111103 (2008).
	
\bibitem{Kuroda_PRB}
    T. Kuroda, S. Sanguinetti, M. Gurioli, K. Watanabe, F. Minami, and N. Koguchi,
    Phys. Rev. B \textbf{66}, 121302 (2002).
    
\bibitem{Kuroda_APL}
	K. Kuroda, T. Kuroda, K. Watanabe, K. Sakoda, N. Koguchi and G. Kido, 
	Appl. Phys. Lett. \textbf{88}, 124101 (2006). 

\bibitem{abbarchi}
	M. Abbarchi, M. Gurioli, S. Sanguinetti, M. Zamfirescu, A. Vinattieri, and N. Koguchi, 
	Phys. Stat. Solidi C \textbf{3}, 3860 (2006)

\bibitem{Kuroda_APEX}
     T. Kuroda, M. Abbarchi, T. Mano, K. Watanabe, M. Yamagiwa,
     K. Kuroda, K. Sakoda, G. Kido, N. Koguchi,
     C. Mastrandrea, L. Cavigli, M. Gurioli, Y. Ogawa, and F. Minami,
     Appl. Phys. Express \textbf{1}, 042001 (2008).

\bibitem{arxiv}
	T. Kuroda, C. Mastrandrea, M. Abbarchi, and M. Gurioli, arXiv:0801.2460.

\bibitem{Loudon}
	R. Loudon, \textit{The Quantum Theory of Light, 3rd ed.} 
	(Oxford University Press, Oxford, 2000) Sec. 8.6.

\end{thebibliography}
\end{document}